# Relation between two proposed Fluctuation Theorems

Denis. J. Evans[*]

Research School of Chemistry, Australian National University, Canberra,

ACT 0200 Australia

Recently van Zon and Cohen [1-3] proposed an extension of the Fluctuation Theorems (FTs) of Evans and Searles [4]. For dissipative nonequilibrium systems, Cohen and van Zon studied the fluctuations of the heat absorbed $Q_t$, over a period of time t, by a surrounding thermostat. They showed theoretically that for thermostatted systems their extension does not exhibit the standard form expected for FTs and $\Pr(\beta Q_t = A)/\Pr(\beta Q_t = -A) \neq \exp[A]$. In the present paper we show that for thermostatted nonequilibrium steady states modeled by Langevin dynamics, the heat function $\beta Q_t$ is in fact identical to the time integral of the phase space compression factor $\beta Q_t = \Lambda_t$ which appears in the Gallavotti-Cohen FT (GCFT). Thus the work of van Zon and Cohen confirms at least for Langevin systems, that the GCFT does not apply to thermostatted steady states.

[*] email: evans@rsc.anu.edu.au

In 1993, Evans, Cohen and Morriss proposed a relation which describes the fluctuation properties of N-particle systems in nonequilibrium steady states that are maintained at constant energy by an appropriate deterministic time reversible *ergostat* [5]. This relation was based on heuristic theoretical arguments, and supported by computer simulation data. Reference [5] motivated a number of papers in which various fluctuation theorems were derived and tested, the first of which were the Evans-Searles Fluctuation Theorems (FTs) [6], and the Gallavotti-Cohen Fluctuation Theorem (GCFT) [7].

Reference [5] considered a very long phase space (steady state) trajectory of a Gaussian ergostatted, (*i.e.* isoenergetic) N-particle system [8]. This long trajectory was divided into (non overlapping) segments of duration t. Along each of the trajectory segments, the instantaneous phase space compression rate, $\Lambda$

$$\Lambda \equiv \frac{\partial}{\partial \mathbf{\Gamma}} \bullet \dot{\mathbf{\Gamma}} \qquad (1)$$

was calculated. Here we denote the phase space vector describing the microstate (coordinates and momenta) of the N-particle system in 3 Cartesian dimensions by $\mathbf{\Gamma} \equiv (\mathbf{q}_1, \mathbf{q}_2, ..\mathbf{q}_N, \mathbf{p}_1, ..\mathbf{p}_N)$. The FT proposed and tested in reference [5] states that for constant energy nonequilibrium steady states:

$$\frac{1}{t} \ln \frac{\Pr(\overline{\Lambda}_t = A)}{\Pr(\overline{\Lambda}_t = -A)} = -A \qquad \text{for large t.} \qquad (2)$$

where Pr(A) denotes the probability of observing A. The standard proofs of the Fluctuation Theorems [4] (FTs) involve the use of time reversible deterministic thermostats to control the temperature or energy [8] of the system of interest. It has been shown that these thermostats can be replaced by a large number of reservoir particles obeying Hamiltonian dynamics. If the heat capacity of the reservoir is sufficiently large, the system of interest can be regarded as effectively isothermal.

In 1994 Evans and Searles derived [6] the first of a set of fluctuation theorems (FTs) for nonequilibrium N-particle systems which focused on a quantity $\Omega$, called the "dissipation rate", rather than on the phase space contraction rate, $\Lambda$ [4]. For thermostatted or ergostatted nonequilibrium steady state systems the long time average "dissipation rate" is identical to the average phase space compression factor and to the average rate of entropy absorption (positive or negative) by the thermostat. To derive an FT one considers an ensemble of trajectories that originate from a known initial distribution (which may be an equilibrium or nonequilibrium distribution, it does not matter) and proceeds under the possible application of external fields and/or thermostats. One then derives general transient fluctuation theorems (FTs) stating that

$$\ln \frac{\Pr(\overline{\Omega}_t = A)}{\Pr(\overline{\Omega}_t = -A)} = At \tag{3}$$

which is of similar form to (2) but where the time averaged phase space contraction rate is replaced by the so-called time averaged dissipation rate or function, $\overline{\Omega}_t(\Gamma)$. In all the transient FTs the time averages are computed from t=0 when the system is characterized by its initial distribution, $f(\Gamma,0)$, to some arbitrary later time t. The dissipation function depends on the initial probability distribution and on the dynamics, and is defined by the equation,

$$\int_0^t ds\, \Omega(\Gamma(s)) \equiv \ln\left(\frac{f(\Gamma(0),0)}{f(\Gamma(t),0)}\right) - \int_0^t \Lambda(\Gamma(s))ds$$

$$\equiv \bar{\Omega}_t t \qquad (4)$$

for all positive times t.

From (1,4) it is trivial to show that if the initial distribution is canonical $f(\Gamma,0) \sim \exp[-\beta H_0(\Gamma)]$ then the phase space compression factor and the dissipation function are related by the equation

$$\Lambda(\Gamma) = \beta \dot{H}_0^{ad}(\Gamma) - \beta \dot{H}_0(\Gamma) \equiv -\beta \mathbf{J}(\Gamma)V\cdot\mathbf{F}_e - \beta \dot{H}_0(\Gamma) = \Omega(\Gamma) - \beta \dot{H}_0(\Gamma) \quad (5)$$

where $\dot{H}_0^{ad}$ is the change in the function $H_0$ caused in the absence of the thermostat. In (5) $\mathbf{J}$ is the dissipative flux, V the system volume and $\mathbf{F}_e$ the dissipative field [8]. For ergostatted dynamics conducted over an ensemble of trajectories which is initially microcanonical, the dissipation function is identical to the phase space compression factor,

$$\Omega(t) = -\Lambda(t) = -[\beta \mathbf{J}](t)V\cdot\mathbf{F}_e, \quad \text{when } dH_0/dt = 0, \qquad (6)$$

where $[\beta \mathbf{J}](\Gamma) \equiv 3N\mathbf{J}(\Gamma)/2K(\Gamma)$, K is the peculiar kinetic energy and $K(0) = 3N\beta^{-1}/2$. However, for thermostatted dynamics (both isokinetic and Nose-Hoover [8]), the dissipation function is subtly different,

$$\Omega(\Gamma) = -\beta \mathbf{J}(\Gamma)V \bullet \mathbf{F}_e, \quad \text{constant temperature}$$
$$\neq \Lambda(\Gamma) = -\beta \mathbf{J}(\Gamma)V \bullet \mathbf{F}_e - \beta \dot{\Phi}(\Gamma) \quad (7)$$

For isokinetic dynamics $3Nk_BT/2 \equiv K, \forall t$, $\beta = 1/k_BT$, $k_B$ is Boltzmann's constant, T is the absolute temperature and $\Phi$ is the potential energy of the system. We note that although the mean of $\dot{\Phi}(\Gamma)$ vanishes for steady states, the difference between the dissipation function and the phase space compression factor cannot be ignored since FTs deal with *fluctuations*. It is clear that for constant temperature dynamics the dissipation function is different from the phase space compression factor. However, in all cases the time averaged dissipation function is equal (with probability one) to the average entropy production since $\lim_{t \to \infty} \overline{[\dot{H}_0]}_t = 0$ and $\lim_{t \to \infty} \overline{[\beta \mathbf{J}]}_t V \bullet \mathbf{F}_e = \lim_{t \to \infty} \overline{\Sigma}_t$ where $\Sigma$ is the extensive entropy production that one would identify for near equilibrium systems from the theory of irreversible thermodynamics [4, 8]. The entropy production is a product of the thermodynamic force $\mathbf{F}_e$ and the time average of its conjugate thermodynamic flux, $\overline{[\beta \mathbf{J}]}_t$.

FTs have been derived for an exceedingly wide variety of ensembles, dynamics and processes [4]. For example FTs have been derived for dissipative isothermal isobaric systems and for relaxing systems where there is no applied external field but where the system is not initially at equilibrium by virtue of its initial distribution $f(\Gamma,0)$. In all cases the FTs have been verified in numerical experiments. Three FTs have recently been confirmed in laboratory experiments: one involving the transient motion of a colloid particle in a moving optical trap [9]; another involving the relaxation of a colloidal particle in an optical trap whose trapping constant is suddenly changed [10] and very recently the time dependent response of an electric circuit [11].

Independently of this activity in 1995 Gallavotti and Cohen [7] proved an asymptotic $t \to \infty$ FT that refers to fluctuations in the phase space compression factor for nonequilibrium steady states, namely (2). This FT is referred to as the Gallavotti Cohen FT (GCFT). They [7] derived the GCFT using the machinery of dynamical systems theory and the Sinai-Ruelle-Bowen measure in particular. A requirement for their derivation to apply is that the system in question should satisfy the Chaotic Hypothesis. The necessary and sufficient conditions for this hypothesis to hold are presently not known but it is thought that the system should be very strongly chaotic and it must obey time reversible dynamics. However it is commonly thought that the GCFT should apply to isothermal steady states as well as constant energy steady states [12].

Shortly after the experimental confirmation of the Evans-Searles FT by Wang et. al. [9], van Zon and Cohen proposed [1,2,3] what they called and "Extended Heat-Fluctuation Theorem". In these papers van Zon & Cohen refer to the dissipation function as the "work function" (thus $\Omega \to \beta W$ in [3]) and the Evans-Searles FTs are referred to as the "conventional FT". For the optical tweezers experiment of Wang et al [9] the proposed extension examined fluctuations in the time average of the quantity

$$\bar{Q}_t = k_B T \bar{\Omega}_t - (1/t)\Delta U(t). \qquad (8)$$

In equation (8) $\Delta U(t)$ is the change in the optical trap potential energy of the particle over the time interval [0,t]. In their paper van Zon and Cohen [3] prove that if the motion of the trapped Brownian particle is governed by Langevin dynamics then in contradistinction to the dissipation function, the heat function fails to satisfy an FT. They show that for transients and steady states, [3],

$$\frac{\Pr(\beta \bar{Q}_t = A)}{\Pr(\beta \bar{Q}_t = -A)} \neq \exp[At]. \qquad (9)$$

[Note: for steady states they show that (9) is not satisfied asymptotically.] However if we look again at the definition of the heat function of van Zon and Cohen (8) and compare it with equation (5) we see that

$$\Lambda(t) = \Omega(t) - \beta \dot{H}_0(t) = \Omega(t) - \beta \dot{\Phi}(t) = \Omega(t) - \beta \dot{U}(t) = \beta Q(t) \qquad (10)$$

For a colloidal system which can be accurately described by the inertialess Langevin equation, $\dot{H}_0 = \dot{\Phi} = \dot{U}$ (the system potential energy is equal to the colloid optical trap potential energy because the solvent is inert). Thus equation (10) shows that for such systems the phase space compression factor is identical to the heat function. This is consistent with the recent demonstration [13] that if heat is lost from a system of interest at a rate Q, to a thermostatting region in thermodynamic equilibrium at temperature T, then the accessed phase space of the system of interest contracts at a rate $\Lambda = Q/k_B T$ [126].

Thus for such a colloidal system the steady state version of the "Extended Fluctuation Theorem" of van Zon and Cohen is in fact the Gallavotti-Cohen Fluctuation Theorem applied to such a system. The fact that this relation fails to describe the fluctuations of a thermostatted colloidal system is in fact consistent with the analysis of Evans, Searles and Rondoni [11] which shows that the GCFT is not applicable to systems which are not maintained at constant energy.

**Acknowledgement**

The author would like to thank Ramses van Zon, Edie Sevick and Debra Searles for useful comments. We would also like to thank the Australian Research Council for partial funding of this work.

**References**

[1] R.van Zon and E.G.D. Cohen, Phys. Rev. Letts., **91**, 110601(2003).

[2] R.van Zon, S. Ciliberto and E.G.D. Cohen, Phys. Rev. Letts., **92**, 130601(2004).

[3] R.van Zon and E.G.D. Cohen, Phys. Rev. E, **69**, 056121(2004).

[4] D. J. Evans and D. J. Searles, Adv. Phys. **51**, 1529 (2002).

[5] D. J. Evans, E. G. D. Cohen, and G. P. Morriss, Phys. Rev. Lett. **71**, 2401 (1993).

[6] D. J. Evans and D. J. Searles, Phys. Rev. E **50**, 1645 (1994).

[7] G. Gallavotti and E.G.D. Cohen, Phys. Rev. Lett., **74**, 2694(1995) and

G. Gallavotti and E.G.D. Cohen, J. Stat. Phys., **80**, 931(1995)

[8] D. J. Evans and G. P. Morriss, *Statistical Mechanics of Nonequilibrium Liquids.* (Academic, London, 1990).

[9] G. M. Wang, E. M. Sevick, E. Mittag, D. J. Searles, and D. J. Evans, Phys. Rev. Lett. **89** (2002).

[10] D. M. Carberry, J. C. Reid, G. M. Wang, E. M. Sevick, D. J. Searles, and D. J. Evans, Phys. Rev. Lett. **92**, 140601 (2004).

[11] N. Garnier and S. Ciliberto, Nonequilibrium fluctuations in a resistor, cond-mat/0407574.

[12] Denis J. Evans, Debra J. Searles, Lamberto Rondoni, On the Application of the Gallavotti-Cohen Fluctuation Relation to Thermostatted Steady States Near Equilibrium, cond-mat/0312353, Phys. Rev. E (submitted).

[13] Stephen R. Williams, Debra J. Searles and Denis J. Evans, Independence of Transient Fluctuation Theorem to Thermostatting Details, Phys. Rev. E. (to appear).